\documentclass[doublecol]{epl2} 

\usepackage{amssymb}
\usepackage{graphicx}
\usepackage{amsmath}
\usepackage{amsfonts}
\usepackage[english]{babel}
\usepackage{wasysym}
\usepackage{units}

\title{Dirac equations in curved space-time versus Papapetrou spinning particles.}
\shorttitle{Dirac versus Papapetrou.} 

\author{F. Cianfrani\inst{1,2} \and G. Montani\inst{1,3,4}}
\shortauthor{F. Cianfrani \etal}

\institute{                    
  \inst{1}ICRA (International Center for Relativistic Astrophysics), c/o Dep. of Physics - ``Sapienza'' University of Rome - Piazzale Aldo Moro 5, 00185, Roma, Italy.\\
  \inst{2} School of Mathematical Sciences, Queen Mary, University of London - London E1 4NS, UK.\\
  \inst{3} ENEA C.R. Frascati - via Enrico Fermi 45, 00044 Frascati, Rome, Italy.\\
  \inst{4} ICRANet - C. C. Pescara, Piazzale della Repubblica, 10 (65100), Pescara, Italy.
}


\abstract{
We find out classical particles, starting from Dirac quantum fields on a curved space-time, by an eikonal approximation and a localization hypothesis for amplitudes. We recover the results by Mathisson-Papapetrou, hence establishing a fundamental correspondence between the coupling of classical and quantum spinning particles with the gravitational field.}

\begin{document}

\maketitle

\section{I- Introduction}

Mathisson \cite{Math} and Papapetrou \cite{1} described the behaviour of a spinning particle in general relativity (GR). The work of Papapetrou is particularly worth noting. He performed a multipole expansion around the worldline. The motion turns out to be non-geodesic, because of the interaction between spin and curvature, while the evolution of the spin consists in a precession around the generalized momentum.\\ 
Since the Mathisson-Papapetrou system does not reproduce the classical force for the spin-orbit interaction \cite{PM2}, Pomeransky and Khriplovich \cite{PM} adopted a different procedure. Given a distribution of energy momentum $T_{\mu\nu}$ in a background metric $g_{\mu\nu}$, they imposed the conservation of the spin tensor along the trajectory and a coupling term of the form $h^{\mu\nu}T_{\mu\nu}$, $h_{\mu\nu}=g_{\mu\nu}-\eta_{\mu\nu}$, $\eta_{\mu\nu}$ being Minkowsky metric.\\ 
The dynamical equations obtained within these two frameworks are different.

On the quantum level, Audrescht \cite{A81} constructed the semi-classical limit of the Dirac equation on a curved space-time using the eikonal approximation (the case with a non-vanishing torsion is presented in \cite{A83}). If the classical 4-momentum was identified with the convection current, the equations dexribing the trajectory and spin evolution turned out to be the Papapetrou equations. However, the transition to the classical picture in terms of a localised worldline was not discussed.\\
Recently, Silenko and Teryaev \cite{ST} found, applying a Foldy-Wouthuysen transformation \cite{FW} on the Hamiltonian, that the equations describing the evolution of the spin and of the momentum associated to a semi-classical spinor in a weak gravitational field coincide with the Pomeransky-Khriplovich equations. 

In this work, the classical-quantum spinning particles comparison is made only when a proper classical description is inferred from quantum theory. This implies that the quantum field must be localized, such that a trajectory can be properly defined. In this respect, we face a second order approach to the semi-classical limit of Dirac particles moving on a curved space-time by an eikonal approximation and a localization hypothesis along the world-line. The latter is obtained to the lowest order of a multi-pole expansion à la Papapetrou (single-pole), such that by spatial integration one deals with quantities having support on the world-line. Within this scheme we are able to derive a dispersion relation from the squared Dirac equation such that general covariance is manifest. No restriction to weak gravitational fields is imposed. Finally we retain the leading order in the semi-classical expansion, which means the first order in the spin, and we get a set of equations which coincide with the Mathisson-Papapetrou ones.\\
For a different approach to recover Papapetrou-like equations from the semi-classical dynamics of Dirac particles, in the context of using Grassmanian variables, see \cite{bcl}.


\section{II- Papapetrou formulation}

In GR a moving body is described by an energy-momentum tensor having support on the world-tube $\tau$. In order to determine the dynamics of the body and of the gravitational field, one should solve the corresponding set of Einstein equations. However, the solution of this system of partial differential equations is a rather difficult task, therefore a very useful approximation consists in neglecting the back-reaction of the body, by assigning a fixed space-time geometry.\\ 
This way the full dynamical information is encoded in the energy-momentum conservation $\nabla_\mu T^{\mu\nu}=0$. Inside $\tau$ one identifies a curve $X^\mu(s)$ and calculates momenta of the body \cite{1}.\\
The multi-pole expansion consists in an expansion around $X^\mu(s)$. At the zero-order one retains only the first moment, which means that the gradients of the gravitational field are neglected over the whole spatial extension of the body. This way, $X^\mu(s)$ results in a geodesic curve, as expected.
At the next order of approximation (the pole-dipole case) gradients of the Christoffel connections are no longer neglected. Hence the following quantity (spin tensor)
\begin{equation}
S^{\mu\nu}=\int_{\tau}\delta x^{\mu}T^{\nu 0}d\tau-\int_{\tau}\delta x^{\nu}T^{\mu 0}d\tau
\end{equation}
is non-vanishing and it enters into equations of motion giving
\begin{equation}
\left\{\begin{array}{c}\frac{D}{Ds}P^{\mu}=\frac{1}{2}R_{\rho\sigma\nu}^{\phantom1\phantom2\phantom3 \mu} S^{\rho\sigma}U^{\nu}\quad\\
\frac{D}{Ds}S^{\mu\nu}=P^{\mu}U^{\nu}-P^{\nu}U^{\mu}\\
P^{\mu}=mU^{\mu}-\frac{DS^{\mu\nu}}{Ds}U_{\nu}\label{pe}\quad\end{array}\right.,
\end{equation}
$m$ and $U^\mu$ being the mass and the 4-velocity of the body, respectively, while $R_{\rho\sigma\nu}^{\phantom1\phantom2\phantom3 \mu}$ is the Riemann tensor.\\
The main conclusions arising in this approach can be summarized as
\begin{itemize}
\item the dynamics is described by the generalized momentum $P^\mu$, which in general is not aligned with the 4-velocity $U^\mu$;
\item the spin tensor motion consists of a precession around the 4-velocity;
\item the trajectory is not a geodesic, since there is an interaction term between the gravitational field and the spin tensor (right side of the first equation in (\ref{pe})).
\end{itemize}
To provide a solution of the system (\ref{pe}) three conditions must be added, because there are thirteen unknown quantities and ten equations only. In the literature one finds three possible choices, 
\begin{equation}
S^{\mu 0}=0,\qquad S^{\mu\nu}U_\nu=0,\qquad S^{\mu\nu}P_\nu=0,\label{concon}
\end{equation}
the Corinaldesi-Papapetrou \cite{2}, the Pirani \cite{Pir} and the Tulczyjew \cite{Tul,Dix} consistency conditions, respectively. 

\section{III- On the localization hypothesis}\label{loc}

The description of a fundamental particle in terms of a classical trajectory around a picked packet is affected by the spread of the wave function, in view of its quantum nature.\\ 
We are going to outline that even though this process occurs, nevertheless a one-particle description can be inferred on sufficiently large macroscopic scales for certain initial conditions.\\  
The key point of our treatment consists in the existence of two length scales. The first one gives the initial wave packet spread and we denote this scale with $\lambda=\alpha\lambda_c$, $\lambda_c$ being the Compton scale, while $\alpha>1$. The second length scale we deal with is the radius of curvature of the space-time manifold, {\i.e.} the one associated with the gravitational field. We indicate it by $L$. In view of applying the theory of quantum fields on a curved space-time, we require $L>>\alpha\lambda_c$. This condition is necessary to recover a proper notion of particle and when violated we enter into the quantum gravity regime.

Given a time-like Killing vector, the energy of the particle can be defined by a dispersion relation of the form $\omega=\sqrt{\frac{\mu^2}{\hbar^2}+g^{ij}k_ik_j}$. 
\\Starting with a wave-packet centered around the wave vector $\vec{k}_0=(k_0,0,0)$, the corresponding wave function reads as follows
\begin{equation}
\phi(x,t)\propto\int d^3k e^{-\lambda^2\frac{(k_1-k_0)^2+k_2^2+k_3^2}{2}}e^{i\vec{k}\cdot\vec{x}-i\omega(k)t},
\end{equation}

$1/\lambda$ being the spread in the momentum space (we take it as isotropic).
As far as a Taylor expansion around $\vec{k}_0$ is concerned, by neglecting terms of the $(k_i-(k_0)_i)^3$ order,  the wave function turns out to be localized around $x=(vct,0,0)$, {\it i.e.} \cite{NM05}
\begin{equation}
\phi(x,t)=\frac{1}{(2\pi)^{3/2}\sigma_p^{1/2}\sigma_t}e^{-\frac{(x^1-vct)^2}{2\sigma_p^2}-\frac{(x^2)^2+(x^3)^2}{2\sigma_t^2}},
\end{equation}
$v=k_0/\omega$. Therefore, the expectation value describes the trajectory of a classical free-particle. However, this is not enough to recover a classical particle dynamics, since the wave packet spreads, thus becoming less and less localized in space. This spread is due to the monotonic increase of deviations $\sigma_p$ and $\sigma_t$ along the directions parallel and orthogonal to the trajectory, respectively.\\ 
Their expressions can be evaluated, finding
\begin{equation}
\sigma_{p/t}(t)=\sqrt{\lambda^2+\lambda_{p/t}\frac{1}{\omega^2\lambda^2}c^2t^2},\qquad \lambda_p=\frac{\mu^2}{\hbar^2\omega^2}\quad\lambda_t=1.
\end{equation}
As a consequence of this spread, soon or later the classical description is no longer available and a trajectory cannot be inferred. The point at which this failure happens depends on the minimum distance the detector can probe and hence can be made arbitrarily large by using a less accurate device.\\
However, our aim is to make a comparison between the behavior of classical and quantum-like spinning objects, as far as their interaction with the gravitational field is concerned. Hence we are interested in a formal definition of a trajectory, which depends only on properties of the beam. We will outline in the Discussion the implications of this choice on the experimental level.\\
For all the speculations above, we take the distance traveled by the expectation value as the characteristic length scale with which the spread has to be compared. Therefore, we set the following condition to yield a
one particle dynamics:
\begin{equation}
vct>>\sigma(t)-\lambda.\label{fun}
\end{equation}      
The behavior of $x^1=vct$ (solid line) and $\sigma(t)-\lambda$ are given by curves in figure 1. 

\begin{figure}[ht]
\onefigure[width=8cm]{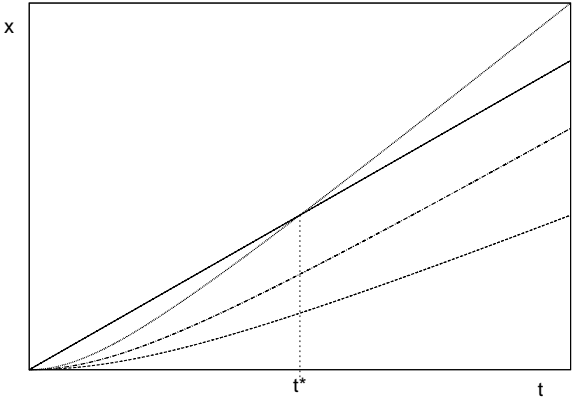}
	\label{fig:mbuto}
	\caption{The solid line gives the expectation value of the position, while the dotted, dash-dotted and dashed ones correspond to the variation of the spread, {\it i.e.} $\sigma(t)-\lambda$, in cases $\alpha v<1$, $\alpha v=1$ and $\alpha v>1$, respectively. It is worth noting how for $\alpha v\geq 1$ the spread always keeps smaller than the proper time along the world-line of the wave packet center.}
\end{figure}

It is worth noting that while for $\alpha v<1$ a one-particle description drops down after a distance $vt^*\approx(\alpha v)^2\lambda<\lambda_c$, for $\alpha v \geq 1$ the spread is always much smaller than the distance. Hence, the localization holds until the quantum gravity regime is approached.\\ 
Therefore, in view of speculations above we deal with wave packets for which 
\begin{equation}
\alpha\geq1/v=\left(1+\frac{\mu^2}{\hbar^2k_0^2}\right)^{1/2}. 
\end{equation}
This condition stands for massless particles, while the initial spread must be larger as the mass increases. However, in view of the great difference between the Compton scale and that one characteristic of macroscopic phenomena, there is a large set of velocities available, without affecting the notion of particle at the initial time.\\ 
We emphasize that this sort of localization allows us to treat particles as being point-like and to neglect spatial gradients in a proper reference frame (see next section) up to the notion of a trajectory is lost. In this approximation scheme the time variation of the spread can be neglected too and in what follows we will treat it as constant.
Indeed this description drops down for observers which are able to test scales much smaller that $L$. 

\section{IV- The semi-classical limit of the Dirac equation}\label{semi}
The Dirac equation in a curved space-time reads as follows (we work in units $\hbar=c=1$ with signature $(+,-,-,-)$)
\begin{equation} 
i\gamma^\mu D_\mu\psi=\mu\psi,\qquad D_\mu\psi=\bigg(\partial_\mu-\frac{i}{2}\omega^{ab}_\mu\Sigma_{ab}\bigg)\psi,\label{direq} 
\end{equation} 
where we have rewritten spinor connections in terms of the gravity spin connections $\omega^{ab}_\mu\equiv e^{\nu b}\nabla_\mu e^a_\nu$ and Lorentz group generators $\Sigma_{ab}=\frac{i}{4}[\gamma_a,\gamma_b]$. If we square the expression (\ref{direq}), by using relation $[D_{\mu},D_\nu]\psi=\frac{1}{8}R_{\rho\sigma\mu\nu}\gamma^\rho\gamma^\sigma$, we obtain 
\begin{equation}
g^{\mu\nu}D_{\mu}D_{\nu}\psi-\frac{1}{4}R\psi=-\mu^2\psi,\label{direq2}
\end{equation}

$R$ being the scalar curvature, {\it i.e.} $R\equiv g^{\mu\nu}R_{\mu\nu}=g^{\mu\nu}g^{\rho\sigma}R_{\mu\rho\nu\sigma}$ and the Ricci tensor $R_{\mu\nu}$ and the Riemann tensor $R_{\mu\nu\rho\sigma}$ retain the expressions as in \cite{LL}.
In what follows, we will address the last equation as the starting point of our dynamical analysis. Although by passing from first- to second-order equations the set of solutions is enlarged, nevertheless in this semi-classical context this procedure has to be regarded as a correct one. In fact, it allows us to treat spinor particles as scalar ones with an $\hbar$ correction due to the spinor structure.

Nevertheless, it is worth noting that differences exist between our method and the one carried on in \cite{ST}, where the semi-classical limit is performed by a Foldy-Wouthuysen transformation.

The semi-classical limit of a quantum field dynamics consists of two steps: the eikonal approximation, by which a classical divergent phase is introduced, and the localization hypothesis for amplitudes, from which a classical trajectory is inferred.\\
The eikonal approximation for spinors reads as follows \cite{Ke02}
\begin{equation}
\psi=e^{iS/\hbar}u,\label{semsp}
\end{equation}
$u$ being a spinor, for which we assume the proper localization along the world-line $x^\mu=X^\mu(s)$, {\it i.e.} we take the product form
\begin{equation}
u(s,x)=\Pi_{\mu=0}^4\frac{1}{\sqrt{2\pi}\sigma_\mu}e^{-\frac{(x^\mu-X^\mu(s))^2}{4\sigma_\mu}}u_0.\label{locspin}
\end{equation}
We take $X^\mu(s)$ as the integral curve of the wave-vector $K_\mu=\partial_\mu S$.\\
Let us now focus on the expression 
\begin{eqnarray}
\int d^3x \sqrt{-g}\bigg[\frac{1}{2}(\bar{\psi}\gamma^{(0)}D^\mu D_\mu\psi+D^\mu D_\mu\bar{\psi}\gamma^{(0)}\psi)-\nonumber\\-\frac{1}{4}R\bar{\psi}\gamma^{(0)}\psi+\mu^2\bar{\psi}\gamma^{(0)}\psi\bigg]=0\label{exp}
\end{eqnarray}
where the Lorentz frame co-moving with $K_\mu$ has been chosen, {\it i.e.} $U_\mu=e^{(0)}_\mu$, $U_\mu$ being the 4-velocity.\\
We point out that the expression (\ref{exp}) vanishes as a consequence of squared Dirac equation (\ref{direq2}).\\ 
Hence, by inserting the form (\ref{semsp}) for $\psi$, $u$ given by (\ref{locspin}), we find
\begin{equation}
(K_\mu K^\mu-K^\mu S_\mu)(1+O(\lambda^2))+\mu^2=0\label{spindis}
\end{equation}
where the $O(\lambda^2)$ corrections come from the expansion of the gaussian around the world-line, while $S_\mu$ reads as 
\begin{equation}
S_\mu=2i\frac{\bar{u}_0\gamma^{\bar{0}}D_\mu u_0-D_\mu\bar{u}_0\gamma^{\bar{0}} u_0}{\bar{u}_0\gamma^{\bar{0}}u_0}.
\end{equation}
Hence the dynamics of $K_\mu$ is obtained by acting on the relation (\ref{spindis}) with the derivative operator $\nabla_\nu$ and we have
\begin{equation}
\left\{\begin{array}{c}U^\mu\nabla_\mu P_\nu-\frac{\hbar}{2}R_{\rho\sigma\mu\nu}U^\mu S^{\rho\sigma}-\hbar\nabla_\nu U^\mu S_\mu-\\-2i\hbar U^\mu D_{[\nu}\bar{u}_0\gamma^{\bar{0}}D_{\mu]}u_0+O(\lambda^2)=0 \\P_\nu=K_\nu-S_\nu\end{array}\right.\label{speqm},
\end{equation}
where $S^{\mu\nu}$ is the charge associated with the spin density of the Dirac field, whose expression reads  
\begin{equation}
S^{\mu\nu}=\frac{\int d^3x\sqrt{h}\bar{u}\{\gamma^{\bar{0}},\Sigma^{\mu\nu}\}u}{2\int d^3x\sqrt{h}\bar{u}\gamma^{\bar{0}} u}=\frac{\bar{u}_0\{\gamma^{\bar{0}},\Sigma^{\mu\nu}\}u_0}{2\bar{u}_0\gamma^{\bar{0}}u_0}+O(\lambda^2).\label{smunu}
\end{equation}
Furthermore, we outline that $S^{\mu\nu}$ couples with the Riemann tensor as the classical spin tensor. Such an identification is enforced by the following relation 
\begin{equation}
S^{\nu\mu}U_\nu=0,\label{Pdir}
\end{equation}
which coincides with Pirani condition (\ref{concon}). 

The hypothesis of dealing with a multi-pole expansion is now translated into a proper condition on $u_0$. We saw in the mono-pole case that the gravitational field can be approximated as being constant over the whole spatial extension of the body. In the same way, here $u_0$ can be treated as a parallel transported spinor on the space orthogonal to $U_\mu$. In this respect, the most general expression one can take for $u_0$ is the following one
\begin{equation}
D_\mu u_{0r}=iU_\mu v_r,\label{u0}
\end{equation}
$v_r$ being an arbitrary spinor, while $r$ is the spinorial index. The condition above is well-grounded by virtue of the analysis of the previous section. We do not expect $u_0$ to appreciate any sensible dependence on spatial coordinates, since in the adopted approximation scheme the full wave function $u$ has a negligible amplitude outside $X^\mu=X^\mu(s)$. On a physical point of view, the particle does not spread to much to ``see'' spatial gradients, but it travels so far to interact with the curvature along the world-line.\\   
This way, equations giving the dynamics can be rewritten as follows
\begin{equation}
\left\{\begin{array}{c} U^\mu\nabla_\mu P_\nu-\frac{\hbar}{2}R_{\rho\sigma\mu\nu}U^\mu S^{\rho\sigma}=0 \\
U^\rho\nabla_\rho S^{\mu\nu}=0
\\ P_\nu=K_\nu-\hbar S_\nu \end{array}\right..\label{dirpap}
\end{equation}
It is worth noting that from the relation $(\ref{u0})$ $P_\mu$ turns out to be aligned with $U_\mu$. Furthermore, being $S_\mu$ proportional to $U_\mu$, the Tulczyjew condition stands as a consequence of the relation (\ref{Pdir}), hence the trajectory is well-defined.\\
In order to determine the dynamics we rewrite $P_\mu=(\mu+n) U_\mu$ and we obtain by the first equation of the system (\ref{dirpap}) 
\begin{equation}
U^\mu\left[\mu\frac{D}{Ds}U_\mu+\frac{D}{Ds}(nU_\mu)\right]=0,
\end{equation} 
which gives $\frac{D}{Ds}n=0$. Hence by a re-definition of the mass, the equation giving the trajectory turns out to be as follows 
\begin{equation}
\left\{\begin{array}{c} U^\mu\nabla_\mu U_\nu-\frac{\hbar}{2}R_{\rho\sigma\mu\nu}U^\mu S^{\rho\sigma}=0 \\
U^\rho\nabla_\rho S^{\mu\nu}=0
\end{array}\right..
\end{equation}

Hence the obtained dynamics is consistent with Papapetrou equations (\ref{pe}) at the lowest order in the spin.

Therefore, Dirac particles follow the trajectory of classical spinning ones, whose spin tensor is given by $S^{\mu\nu}$ (\ref{smunu}). An interaction with the gravitational field is predicted. We emphasize that our results differ with respect to \cite{Mum}, where equations of motion are inferred from a semi-classical Lagrangian density.
 
The whole approximation scheme is consistent, since spinors can ``see'' only those components of the Riemann tensor with at least one time-like index, {\it i.e.} those ones which do not contribute to the tidal forces on the spatial hypersurfaces. Such a picture has been suggested by results of the previous section. We have seen that the one-particle approximation stands as soon as $\alpha v\gg 1$, while quantum gravity enters into the physical description for $t\sim \alpha v L$. Therefore, we expect that $\alpha$ can fixed such that the particle has enough time to interact with the geometry along its trajectory, while the space on which the packet spreads remains effectively flat. 

For instance, the analogous conclusion can be inferred from the classical analysis of the trajectory in a Schwarzschild space-time. In this respect, in \cite{Pl98} it is outlined that the deviation from a geodesic motion is due to the interaction of the spin with the gravito-magnetic part of the metric, while tidal forces do not affect the dynamics.

Finally, we want to stress that the assumptions made on the wave function (\ref{semsp}) ({\it i.e.} the eikonal approximation and the localization hypothesis) are crucial to infer the above result. Hence the application of the Ehrenfest theorem would not be enough to obtain a Mathisson-Papapetrou trajectory.

\section{V- Discussion}
We outlined in section III that our results allow to make a formal comparison between the trajectories of quantum and classical spinning objects. Even though in principle experiments can be realized to probe such a model, the magnitude of the predicted effects is too weak to be detected.\\ 
The main consequence of the picture outlined is that the particle trajectory deviates from geodesic motion. This deviation into a Schwarzschild space-time is discussed in \cite{PSK00} (see formula (35)), where the Pomeransky-Kriplovich and the Mathisson-Papapetrou results are compared. The main conclusion of that analysis is that different deviations are predicted and only in the first case the classical spin-orbit interaction is reproduced.\\ 
However, for Earth-based experiments the tidal forces are so weak that the associated geodesic deviation takes place on length scales remaining within the dimension of the packet. As far as particles coming from extra-galactic sources are concerned, even though we are supposed to deal with beams having a proper energy (no pair production), the semiclassical description is not more available for observers on the Earth. Another possible experimental confirmation could come from COW-like \cite{Cow1,Cow2} experiments. Since the spin couples to the curvature, trajectories at different height are characterized by a different time of flight, which add a further contribution to the quantum interference. However, such an additional term is well below experimental uncertainty \cite{Cow3}.\\    
A different way of detecting the interaction between the spin and the gravitation field is the appearance of an effective polarization. Given a source emitting a beam of Dirac particles whose polarization oscillates with a certain frequency $\Omega$, the different time of flight for particles with different spins gives a correction to the polarization measured by a distant observer. For a beam moving in a Kerr space-time along a trajectory on the equatorial plane, we find
\begin{equation}
\frac{\delta\omega}{\omega}\approx\frac{3\omega L^2Ma\lambda_c}{2vr^4}.
\end{equation}    

$M$ and $a$ are the two parameters of the Kerr solution, the mass and the angular momentum per unit of mass, respectively, while $r$ is the radius of the trajectory, and $L$ is the distance traveled. For the Earth one gets 
\begin{equation}
\frac{\delta\omega}{\omega}\approx\omega L^2\cdot10^{-47}\frac{s}{m^2} 
\end{equation} 

However, the realization of experiments on polarized beams of Dirac particles that are capable of testing such small deviations is rather remote (for photons similar deviations have been confirmed experimentally, see \cite{ST07} and references therein). 
   
\section{VI- Concluding remarks}
In this work we have focused our attention on the semi-classical limit of the Dirac equation in a curved space-time. The main point is the way this limit has been taken. While the eikonal approximation is the standard tool to derive  classical mechanics from quantum mechanics, the hypothesis of localization is proper of our treatment, since other approaches do not consider it \cite{A81,NSH92}. But we want to stress that to make a comparison with the behavior of classical objects, a classical trajectory must come out from the quantum description.\\ 
We obtain such a feature by a multi-pole expansion and, since we are describing an intrinsic spin, we have to retain only the monopole term when taking the classical limit. The best we can do is to localize wave functions at the Compton scale order, but we were able to define quantities on their mean values by a spatial integration. Such a kind of integration is proper of the Papapetrou approach and this way a trajectory is well-defined.
 
This formulation provided with a notion of spin tensor, whose quantum character is manifest, because it arises as a first order correction in the semi-classical approximation, and whose physical interpretation is clear as the expectation value of the spin density. Furthermore, general covariance is manifest during all calculations, while no weak-field approximation takes place for the gravitational field.

The dynamics reproduces the results by Mathisson and Papapetrou at the first order in the spin. Hence, there is no agreement with the papers \cite{ST}, where the semi-classical limit is performed by a Foldy-Wouthuysen transformation on the Hamiltonian and the Pomeransky-Kriplovich formulation is obtained. These two different procedures allow us to infer a semi-classical picture, but they provide different results for the classical dynamics.

Therefore, it is possible to infer from a quantum description both the two compelling formulations for spinning particles in GR. Only the experimental detection of deviations from geodesic motion would tell us what is the appropriate semi-classical procedure for Dirac fields in curved space-time. 
 

\section{Acknowledgment} 
We thank Claus Laemmerzahl for having called our attention on \cite{A81,A83}.

\end{document}